\documentclass{article}
\usepackage{graphicx} 

\newcommand\blfootnote[1]{%
  \begingroup
  \renewcommand\thefootnote{}\footnote{#1}%
  \addtocounter{footnote}{-1}%
  \endgroup
}
\usepackage{enumerate}
\usepackage{enumitem} 
\usepackage{authblk}
\usepackage[a4paper, margin = 1in]{geometry}
\usepackage{booktabs}
\usepackage[sort&compress,numbers]{natbib}

\title{Asynchronous Training of Mixed-Role Human Actors  in a  Partially-Observable Environment}

\author[1]{Kimberlee Chestnut Chang\footnote{These authors contributed equally to this research.}}
\author[1]{Reed Jensen$^*$}
\author[1]{Rohan Paleja$^*$}
\author[1]{Sam L. Polk$^*$\footnote{Corresponding author; email: samuel.polk@ll.mit.edu.}}
\author[1]{\newline Rob Seater}
\author[1]{Jackson Steilberg}
\author[1]{Curran Schiefelbein}
\author[2]{Melissa Scheldrup}
\author[3]{\newline Matthew Gombolay}
\author[1]{Mabel D. Ramirez}

\affil[1]{M.I.T. Lincoln Laboratory, Lexington, MA, USA}
\affil[2]{MetaTeq Inc., Eugene, OR, USA}
\affil[3]{Georgia Institute of Technology, Atlanta, GA, USA}

\date{}

\begin{document}

\maketitle

\begin{abstract}

    In cooperative training, humans within a team coordinate on complex tasks, building mental models of their teammates and learning to adapt to teammates' actions in real-time.
    To reduce the often prohibitive scheduling constraints associated with cooperative training, this article introduces a paradigm for \textit{cooperative asynchronous training of human teams} in which trainees practice coordination with autonomous teammates rather than humans.  
    We introduce a novel experimental design for evaluating autonomous teammates for use as training partners in cooperative training. 
    We apply the design to a human-subjects experiment where humans are trained with either another human or an autonomous teammate and are evaluated with a new human subject in a new, partially observable, cooperative game developed for this study.
    Importantly, we employ a method to cluster teammate trajectories from demonstrations performed in the experiment to form a smaller number of training conditions. This results in a simpler experiment design that enabled us to conduct a complex cooperative training human-subjects study in a reasonable amount of time. 
    Through a demonstration of the proposed experimental design, we provide takeaways and design recommendations for future research in the development of cooperative asynchronous training systems utilizing robot surrogates for human teammates. 
\end{abstract}

\blfootnote{\newline \noindent DISTRIBUTION STATEMENT A. Approved for public release. Distribution is unlimited.

\noindent This material is based upon work supported by the Under Secretary of Defense for Research and Engineering under Air Force Contract No. FA8702-15-D-0001. Any opinions, findings, conclusions or recommendations expressed in this material are those of the author(s) and do not necessarily reflect the views of the Under Secretary of Defense for Research and Engineering.} 

\section{Introduction}
Asynchronous training paradigms mitigate scheduling constraints of teachers and students by enabling easier access to training material, typically through computer-based self-instruction~\cite{gerencser2020review, marano2020review, serna2016teaching}. 
While this paradigm has been successful in \textit{teaching an individual} across several high-impact domains---e.g., in medicine~\cite{gaete2023remote, pourmand2018impact} and social work~\cite{gerencser2020review}---using asynchronous training to \textit{teach a team} to coordinate on complex tasks is understudied and has yet to be shown as effective. 

In several applications including game-playing (e.g., a Dota 2 team), disaster response (e.g., a team of firefighters), and sports (e.g., a soccer team), it is critical for team members to not only maintain individual proficiency, but also to collaborate fluently.  
However, synchronous training (where all teammates are present) is often costly due to increased time requirements and associated scheduling constraints. The creation of cooperative synchronous training systems is further complicated in settings where teammates have different roles, encounter partial observability (i.e., essential information may be siloed across teammates), or cannot directly communicate due to tight time constraints or a lack of communication links. 
In such settings, however, it is not only essential for each trainee to be proficient in their own role but also to learn mental models for all teammates to enable adaptation to their stylistic preferences and strategy in evaluation~\cite{andrews2023role}. Indeed, the ability of teammates to build accurate shared mental models~\cite{andrews2023role, endsley2000situation} of the environment and each other has been shown to be very effective in team-building and team situational awareness in partially-observable environments~\cite{bolstad1999shared, Paleja2020InterpretableAP, Paleja2022TheUO, andrews2023role, endsley2000situation}.  

Additional complications with evaluating synchronous training paradigms are: 1) true appraisal of these systems via human-subject studies can prove challenging, as it often requires multiple days to train users to develop effective individual and collaborative strategies;  
2) it can be difficult to tease out changes to individual and team performance in tight collaboration settings;  
3) it is not straightforward to understand how well subjects' training extends to novel, unfamiliar teammates.  
To reduce the need for costly human-human training and address related concerns, 
we must develop \emph{cooperative asynchronous training} paradigms that help users build mental models of their expected (or unexpected) teammates, utilizing autonomous robot surrogates for human teammates that approximate their behavior and strategies.

This paper introduces a generalizable paradigm for evaluating the efficacy of an autonomous agent as a surrogate training partner to human trainees in a cooperative asynchronous training paradigm.
To focus our analysis on a challenging team collaboration paradigm with strict roles and partial observability, we first introduce a teaming game titled \textit{Overcooked-AI: Have You Been Served?} (HYBS) with two complementary roles, each having access to different information. The game design allows for variation in individual strategies and rewards cohesive coordination. Second, we present a novel experimental design that utilizes user strategy preferences to evaluate an autonomous agent's efficacy as a part of cooperative asynchronous training paradigm. In our human-subjects study, we use four different training paradigms, including a) concurrent human-human training (the costly standard today), b) cooperative asynchronous training with an imitation-learning-based agent trained on human data, c) cooperative asynchronous training with a heuristic-based, pre-programmed agent, and d) no training (i.e., a baseline condition). We employ a combination of objective and subjective measures to understand a human's ability to learn to collaborate both with a teammate who executes a strategy similar to what was observed in training (in-sample performance) and with a teammate that employs a different strategy (out-of-sample performance)~\cite{polk2024unsupervised}.

This article makes three significant contributions. 
First, we introduce a novel experimental design for evaluating the efficacy of autonomous agents as training partners in a partially observable environment with mixed roles. 
We find that segmenting our human-subject research design by behaviors learned through sequential clustering of demonstrator decision trajectories~\cite{polk2024unsupervised} rather than by individual demonstrators drastically reduces the number of conditions (and hence participants) required for evaluation of asynchronous training outcomes. 
Second, although we did not find that the training condition significantly affected evaluation outcomes in an implementation of this experimental design in a real-world human-subjects experiment, we provide important guidelines for future research on the development of cooperative asynchronous training systems and associated autonomous training partners. 
Finally, we provide results on human perception of autonomous teammates from our demonstration of the proposed experimental design. 
Indeed, despite significant differences between the evaluated autonomous teammates' objective performance, we find that humans that teamed with either agent teammate were rated equally poorly compared to human teammates, suggesting that humans may rate based on teammate behavior in addition to performance.

The rest of this article is structured as follows. In Section \ref{sec: Related Work}, we overview related work on computer-assisted training systems and human modeling. 
In Section \ref{sec: Overcooked overview}, we introduce HYBS. 
In Section \ref{sec: experimental_design}, we overview the teammate generation procedure in our novel, proposed experimental design, which is described in detail in Section \ref{sec: human-subjects design}.
In Section \ref{sec: human_subject_exp}, we analyze key metrics on training outcomes produced from a real-world human-subjects experiments demonstrating the proposed experimental design,  along with design recommendations for future research on cooperative asynchronous training. 
We conclude in Section \ref{sec: Conclusion}.

\section{Related Work} \label{sec: Related Work} 

\subsection{Asynchronous Training Systems} \label{sec: Computer-Assisted Training}
Single-agent tasks (e.g., learning algebra) have been supported through computer-assisted instruction systems which  guide users through customized lesson plans adapted to students via personalized feedback~\cite{Carbonell1970AIIC, marano2020review, gerencser2020review, serna2016teaching, gaete2023remote, pourmand2018impact, goel2018jill}. These asynchronous systems are beneficial in reducing the time requirements for expert teachers but are often expensive to design, requiring domain expertise from subject matter experts, education experts, and software developers. More recent research has employed solutions to take advantage of crowdsourced data~\cite{moubayed2018learning, johnson2016potential, liu2017using},  build autonomous robot teaching assistants to human instructors~\cite{ goel2018jill, abendschein2021human}, and learn adaptive teaching systems based on trial and error with real students~\cite{Spain2021ARL}. Notably, when evaluating these approaches using simulated human models as students, results can widely differ from those obtained through a real-world human-subjects study, underscoring the importance of proper evaluation~\cite{Zook2015TemporalGC}. 

\emph{Cooperative asynchronous training systems} extend these systems to multi-agent coordination problems  (e.g., teaming in partially observable domains such as StarCraft II and Dota 2), where the end-goal is to play well on a human-only team. A cooperative asynchronous training system should teach both individual proficiency as well as team fluency. In these domains, automated tutorials are often hand-crafted by game developers and are meant to provide only individual proficiency, not collaborative fluency. 
\textbf{To the best of our knowledge, this article presents the first evaluation of a cooperative asynchronous training system in a partially observable, multi-agent setting.}

\subsection{Human Modeling} \label{sec: human model} 

The successful development of asynchronous cooperative training using autonomous agents requires representative human surrogate models to serve as training partners. Several techniques exist for generating a model of human behavior, ranging from rule-based \emph{heuristic} models~\cite{orkin2004goap}, based on psychology principles and domain expertise, to data-based techniques which leverage collected data to infer an \emph{apprentice} model of user behavior~\cite{Li2017InfoGAILII,Paleja2020InterpretableAP,Abbeel2004ApprenticeshipLV}. Data-based techniques are typically able to capture more accurate and personalized models of user behavior~\cite{Paleja2020InterpretableAP, andrews2023role, Paleja2022TheUO, wang2019learning}, but often suffer in low-data regimes and agent behavior may not always perform well in all regions of the state space. Heuristic, pre-programmed behavior is typically more robust and predictable~\cite{orkin2004goap, siu2021evaluation}, but the extent of personalization is limited and the quality of the heuristic is dependent on domain expertise. 
Data-based human models therefore may have uses in domains where manual hand-crafting of heuristic agents by experts is infeasible but where ample demonstration data exists for inference~\cite{wang2019learning}.
We include both a heuristic and apprentice model as training partners in our human-subjects study, presenting key takeaways regarding these models as surrogate training partners.

\section{Overcooked-AI: Have You Been Served?} \label{sec: Overcooked overview}

Overcooked is a cooperative resource management and coordination game where chefs coordinate to prepare and serve appropriate dishes to customers. Chefs must navigate obstacles, collect and prepare ingredients, cook items, and serve prepared dishes. Players are rewarded with tips from customers for fulfilling requested orders in a timely manner. A simplified version---Overcooked-AI~\cite{carroll2019utility}---has been developed as a research testbed for evaluating teams of human- and autonomously-controlled chefs in a fully-observable setting. 

\begin{figure}[t]
    \centering
    \includegraphics[width = 0.65\textwidth]{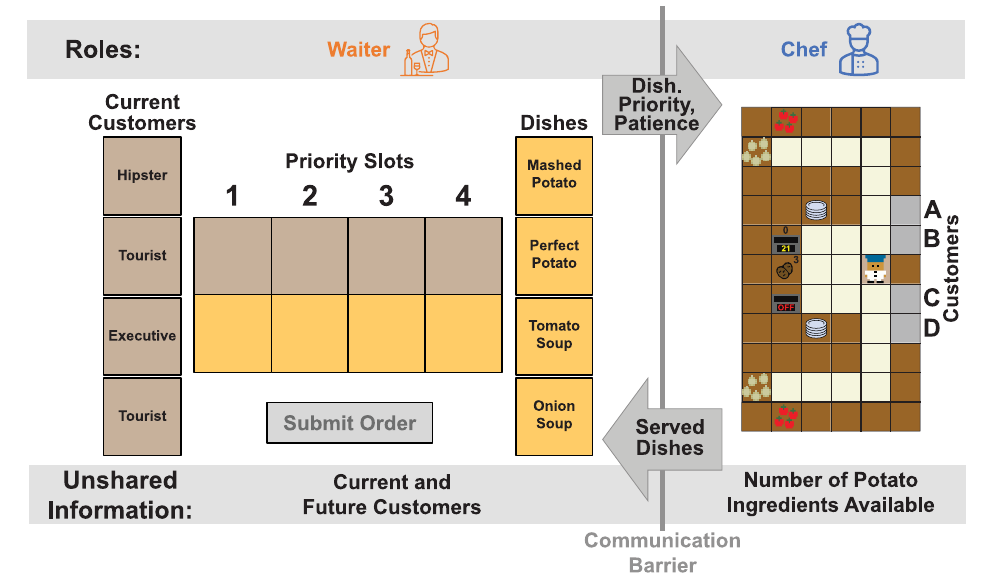}
    \caption{
    Visualization of gameplay in the HYBS partially-observable multi-agent teaming game. 
    The novel waiter role assigns a priority to each customer and recommends a dish for them to be served, which the chef role uses to plan and execute dish servings in a round. 
    The waiter and chef have different sets of information and must coordinate in pursuit of tip rewards by serving dishes to customers matching those customers' preferences.
    }
    \label{fig: Overcooked}

\end{figure}

This article focuses on a partially observable teaming setting in which outcomes depend on coordination between users with different roles and access to different pieces of essential information. 
To this end, we introduce HYBS: a version of Overcooked-AI with several key changes, including a new ``waiter'' role. The waiter is tasked with assigning dishes to customers (who have different dish preferences) and ordering customers within a priority queue without knowing ingredient amounts remaining in the kitchen. The chef, in turn, has no access to customer preferences or future customer arrivals and must rely on the waiter's recommendations. A visualization of our environment is displayed in Fig. \ref{fig: Overcooked}. Other key changes include using a movement-based clock to emphasize the importance of strategy and the addition of a ``potato'' ingredient: a highly-effective but limited-inventory resource.

A game in HYBS consists of seven turn-based rounds, starting with one short chef preparation turn, during which, the chef can stage ingredients by placing them closer to stoves in anticipation of upcoming demand. The waiter and chef then alternate gameplay three times in \emph{performance rounds}, teaming to fulfill customer orders.   
In waiter performance rounds, the waiter receives information (dish preferences and tip ranges) on the four customers to be served in the next chef performance round and a description of customer profiles to be served in later rounds. The waiter's role is to recommend dishes for the chef to serve to each of four customers in the next chef performance round and the order in which to serve them for maximal reward. Notably, this task is not trivial as the waiter is unaware of potato resource availability, nor do they see which ingredients may have been previously staged.

In chef performance rounds, the chef must gather ingredients, prepare dishes, and serve customers based on the waiter's recommendations, subject to resource and time constraints. To complete a dish, the chef must collect ingredients stationed around the kitchen and place them within pots to begin cooking. To serve a cooked dish, the chef plates the dish and carries it to the customer for an immediate reward, if delivered before the customer leaves and becomes unservable. Chefs may also stage ingredients and dishes on countertops in the scene for later retrieval within the current (or a future) performance round. 
Importantly, the chef is aware of potato resource availability, but the waiter is not, which may motivate deviating from waiter dish assignments. Unlike the waiter, the chef does not know the customer profiles associated with service recommendations or those of future performance rounds. The limited observability across the team makes it important for players to  build a mental model of their teammate through training that accounts for stylistic preferences~\cite{andrews2023role}. 
We provide further information in this article's supplement, alongside additional figures/videos and specific details about game initialization and dynamics.

\section{Teammate Generation For Cooperative Asynchronous Training Study}  \label{sec: experimental_design}

A key component of our experimental design is the creation of confederates to serve as training and evaluation partners.
With relation to our domain of HYBS, the proposed experiment is concerned with training waiters to cooperate with chefs. Thus, we need a population of established chef players that can serve as training partners to new waiters, as well as chef players that can serve as never-before-seen evaluation partners with trained waiters. 
This section overviews the important step of generating these chef teammates for the proposed cooperative asynchronous training experiment, which will be described  in Section \ref{sec: human-subjects design}.

\subsection{Established Chef Generation and Clustering} \label{sec: Human Chefs}

A set of participants (i.e., \emph{established chefs}) is selected to train by playing both waiter and chef roles across ten games\footnote{Based on a pilot study, we identify ten games as generally sufficient for players to establish specific strategies that lead to maximizing cumulative tip.} of HYBS (i.e., self-play). These participants are selected based on availability, as established chefs need to return on a different day to train and evaluate with waiters during the main portion of the experiment.
We note that, although utilizing self-play enables the training of human chef players, the trained players and data obtained are not a mirror of the actual game setting, as the user has complete information. This may have downstream effects (e.g., distribution shift) in deploying apprentice models trained on established chef demonstrations with full observability.

To understand the diversity in established chef behaviors and pair users properly in training and evaluation, we must identify common behaviors exhibited by established chefs during training.
Behaviors among the fourteen chefs in the established chef population are located through training an unsupervised decoder-free, sequential encoder (the architecture class proposed in~\cite{franceschi2019convolutionalencoder}) to map each game's associated sequence of state-action pairs into a two-dimensional representation space in which games can easily be compared and analyzed~\cite{polk2024unsupervised}. The state-action pairs used consist of carefully-selected state variables meant to be explanatory of the game at a given time and the action that the player took given that status of the game; see our supplement for details. Sequences of state-action pairs were partitioned into train (70\%) and validation (30\%) sets, and we selected the sequential encoder among a grid of trained encoder with lowest encoding loss on the validation set after training. The resulting sequential encoder is highly successful at encoding sequences of state-action pairs, with a training loss of 1.05 and validation loss on 1.13; the loss function used is the one proposed with the implemented sequential encoder~\cite{franceschi2019convolutionalencoder}.

\begin{figure}[b]
    \centering
    \includegraphics[width = 0.51\textwidth]{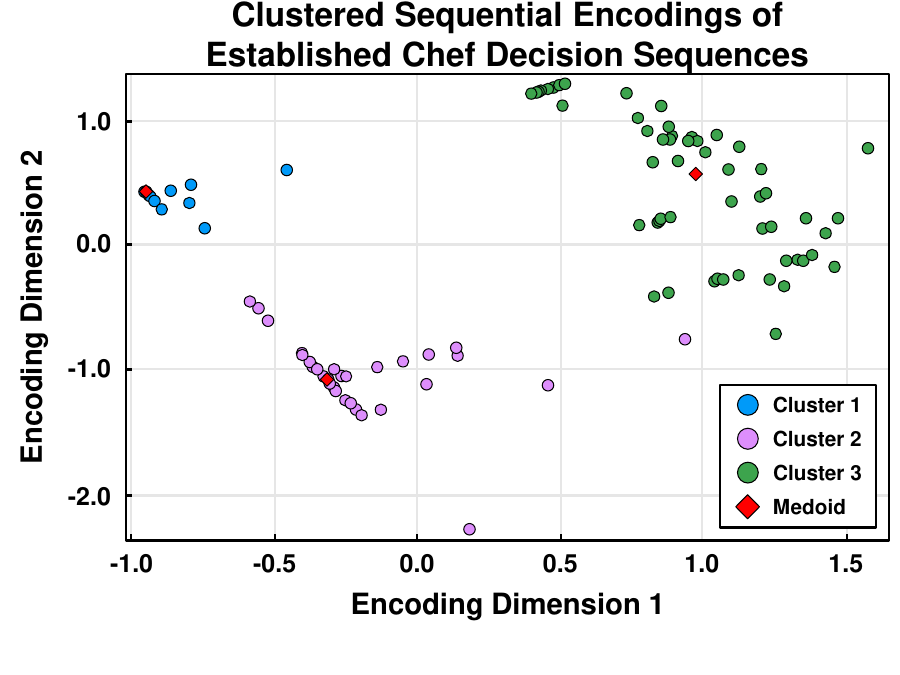}
    \caption{Behavior clustering obtained from $K$-Medoids applied to sequential encodings of established chef demonstrations.  
    }
    \label{fig: UNBIAS clustering}

\end{figure}

After obtaining a behavior representation through sequential encoding, we employ $K$-Medoids~\cite{hastie2009elements} across $K=2,3,4,5$ clusters. The clustering maximizing silhouette score~\cite{rousseeuw1987silhouettes} (visualized in Fig. \ref{fig: UNBIAS clustering}) is used to associate chef gameplay with behaviors~\cite{polk2024unsupervised}. To further validate that the obtained clustering relates to distinct behaviors, we consider the distribution of each subject's gameplay across clusters.  For each of thirteen out of fourteen established chefs, at least half of that subject's gameplay is concentrated in a single cluster, as indicated in Table \ref{tab:matching matrix}. Notably, the one subject for which this is not true---chef ``2''---stated that they attempted to vary their strategy across the games played, possibly explaining the uniformity of their gameplay across cluster assignments.  Nevertheless, because our unsupervised sequential clustering procedure was naive to subject identity, the overlap of the resulting cluster assignments with subject identity indicates the unsupervised learning of blocks of commonly-expressed human behaviors, as desired.

Cluster-to-chef assignments enable \textit{in-sample} evaluation of a trained waiter (by playing with a different established chef from the same cluster as the chef the waiter trained with) as well as \textit{out-of-sample} evaluation (by playing with an established chef from a different cluster from the chef the waiter trained with). 
Without these cluster assignments, testing with an out-of-sample established chef would require evaluation with all waiter-established chef pairs. 
Thus, behavior clustering of established chefs ultimately plays a key role in the creation of a feasible user study.

\begin{table}[t]
    \centering
    \scriptsize
    \begin{tabular}{ccccccccccccccc}
        \toprule
        Chef ID  & 1            &   2           &   3           &   4           &   5           &   6           &   7           &   8           &   9           &   10          &   11          &   12          &   13          &   14          \\
        \midrule 
        Cluster 1& 2            &   3           &   1           &   3           &   \textbf{5}  &   1           &   0           &   4           &   1           &   3           &   1           &   2           &   0           &   1           \\
        Cluster 2& 2            &   3           &   3           &   2           &   3           &   \textbf{6}  &   2           &   0           &   3           &   \textbf{7}  &   \textbf{6}  &   \textbf{7}  &   1           &   \textbf{7}  \\
        Cluster 3& \textbf{6}   &   \textbf{4}  &   \textbf{6}  &   \textbf{5}  &   2           &   3           &   \textbf{8}  &   \textbf{6}  &   \textbf{6}  &   0           &   3           &   1           &   \textbf{9}  &   2           \\ 
        \bottomrule
    \end{tabular}
    \caption{Matching matrix associating unsupervised clustering of encoded sequences of state-action pairs from established chefs with their ID labels. Values indicate the number of games played by a subject assigned to each cluster. All subjects except for chef 2 had at least half of their gameplay concentrated in a single cluster (bolded). Because the sequential clustering procedure used is naive to chef ID labels, this overlap indicates learning of commonly expressed behaviors in HYBS.}
    \label{tab:matching matrix}
\end{table}
 
\subsection{Agent Chef Generation} \label{sec: Agent Chefs}

We develop two agent chef models for the proposed asynchronous training paradigm: the \emph{apprentice} and \emph{heuristic}. Importantly, each agent is intended to represent a class of generation techniques (e.g., the apprentice is tied to imitation-learning techniques) and may be associated with several different behavior clusters. We provide high-level overviews of each agent in this section and provide more detail in this article's supplement.

At the beginning of each turn, each agent identifies the customer-dish pairs that should be created based on waiter recommendations. The chef executes customer-dish pairs using a combination of Monte Carlo Tree Search (MCTS)~\cite{kocsis2006bandit} for planning high-level actions (e.g., get onion, put onion in pot to cook, and serve customer onion soup) and a Goal-Oriented Action Planner (GOAP)~\cite{orkin2004goap} for planning and executing low-level actions (e.g., move, place, and wait). As such, the main difference between the two agents is how customer-dish pairs are selected. 

The \emph{apprentice} agent chef is trained on chef trajectories collected during training of established chefs in Section \ref{sec: Human Chefs}. This model learns a supervised mapping from waiter recommendations and the current state of HYBS to a customer-dish pair. The aforementioned MCTS-GOAP planner carries out lower-level gameplay, given this assignment. To train a model to mimic the established chefs, we follow the learning algorithm of \cite{Paleja2020InterpretableAP}, which augments state-action pairs with a latent personalized embedding that represents a human's behavior. These latent embeddings are initialized uniformly for each chef, and learned through training. Specifically, during training, we repeatedly sample data from a random user, load in that user's personalized embedding, and follow the following procedure: 1) update the apprentice model parameters and personalized embedding via a cross-entropy loss that maximizes imitation performance; 2) update all parameters utilizing a mutual-information maximization loss that maximizes the relation between sampled embeddings and state-goal pairs (i.e., mode discovery).
After training this mode, in an end-to-end fashion, simultaneously learning a policy of behavior and personalized embeddings that represent users, personalized embeddings can be used to condition the apprentice and generate stylistic preferences observed in human customer-dish assignments. 
In this study, personalized embeddings are learned from behavior clusters from Section \ref{sec: Human Chefs} (rather than from individual demonstrators) so that apprentice models imitate the behaviors inferred from established chef self-play.

The \emph{heuristic} agent chef is programmed to be obedient to the waiter's recommendations and uses MCTS to discover a valid set of high-level actions that fulfill the recommendations. The agent then uses GOAP to plan low-level actions. If, at any point, it finds that the waiter's goals are not achievable, then it will attempt to follow a feasible subset. 
If time remains at the end of the round, the heuristic is programmed to stage ingredients for future rounds (i.e., place ingredients on a closer counter to reduce the number of steps for retrieval in future rounds). We implemented two variants of the heuristic, wherein either tomatoes or onions were staged for later use.

\section{Human-Subjects Study Design} \label{sec: human-subjects design}
\begin{figure*}[t]
    \centering
    \includegraphics[width = 0.99\textwidth]{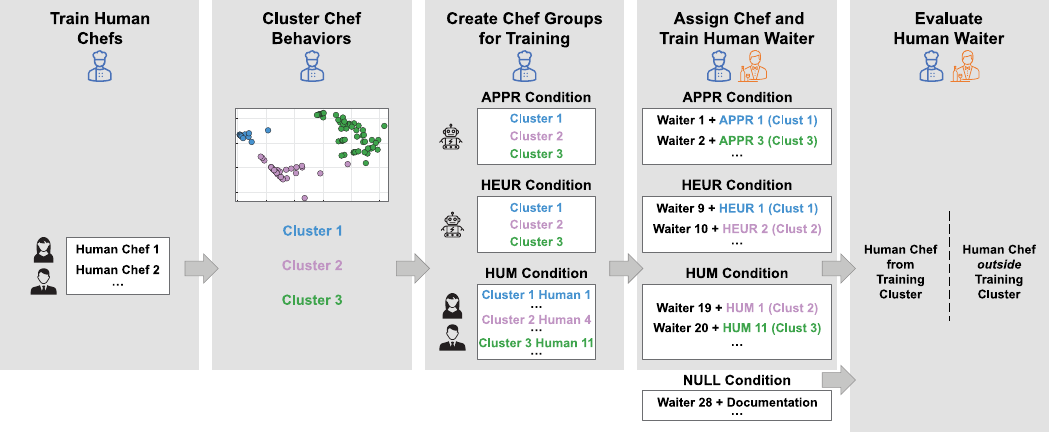}
    \caption{Overview of proposed asynchronous cooperative training experiment. Established chefs train via self-play, producing data used to learn clusters of commonly-expressed chef behaviors and train agent chefs. Waiters are randomly assigned a training partner---a human or agent (apprentice or heuristic) chef---and cluster. A control condition receives no training treatment. All chefs are evaluated with an established human chef from the cluster they trained with and one from an unseen cluster. }
    \label{fig:Experiment}

\end{figure*}

Finally, the experiment relies on a human-subjects study design for training in an asynchronous cooperative setting in HYBS (visualized in Fig. \ref{fig:Experiment}).

\vspace{0.05in}

\noindent\textbf{Independent Variables:}
We have one independent variable in our experiment, the training partner with whom the waiter is paired. Training conditions are not revealed to the user (e.g., a waiter training with a human chef is not told they are teaming with a human). We consider the following conditions:
\begin{enumerate}
    \item \textbf{(HUM) Waiter Training with Human Chef:} 
    The waiter is paired with an established human chef.
    \item \textbf{(APPR) Waiter Training with Apprentice Chef:} 
    The waiter is paired with an apprentice chef.
    \item \textbf{(HEUR) Waiter Training with Heuristic Chef:} 
    The waiter is paired with a heuristic chef.
    \item \textbf{(NULL) Control:} 
    The waiter receives no training.
\end{enumerate}
Each participant serving as waiter is randomly assigned to a specific training condition chef. The chef will be associated with one of three  behaviors, determined \textit{a priori} via clustering in Section \ref{sec: Human Chefs}. This information assists in measuring the how subjects ability to work with teammates exhibiting similar behaviors to that which was learned during training (in-sample performance) translates to the ability to work with teammates exhibiting a very different strategy (out-of-sample performance).  
In the \textbf{APPR} condition, a personalized embedding inferred during training is used to condition the agent to present stylistic behavior associated with the assigned cluster. 
In the \textbf{HEUR} condition, Cluster 1 and Cluster 2 participants are assigned the heuristic variant associated with common staging behaviors within those clusters (preferencing either tomato or onion ingredients), as these were the two ingredients that may be staged in HYBS. \textbf{HEUR} participants assigned to Cluster 3 trained with a random variant among the two implemented heuristics.
Waiters assigned to the \textbf{HUM} condition were assigned chefs from the established chef pool, which were each assigned a behavior cluster (see Section \ref{sec: Human Chefs}). We did not enforce established chefs' compliance with their previous behavior. 

\vspace{0.05in}

\noindent\textbf{Research Questions:}
We look to address the following research questions by comparing across different conditions:

\begin{enumerate}
    \item[\textbf{Q1}] \textbf{Effectiveness}: Does training with an autonomous teammate yield evaluation outcomes comparable or superior to training with a human? Addressing this question is accomplished by comparing the \textbf{APPR} and \textbf{HEUR} conditions to \textbf{HUM}.
    \item[\textbf{Q2}] \textbf{Personalization}: Does training with an autonomous teammate that explicitly encodes personalization and responds to waiters' heterogeneous strategies result in evaluation outcomes superior to training with one that does not?  Addressing this question is accomplished by comparing the \textbf{APPR} condition to  \textbf{HEUR}.
    \item[\textbf{Q3}] \textbf{Benefit}: Does training result in evaluation outcomes superior to not training? This can be assessed by comparing the \textbf{APPR}, \textbf{HEUR}, and \textbf{HUM} conditions to \textbf{NULL}.
    \item[\textbf{Q4}] \textbf{Perception}: Does training with an autonomous teammate lend a similar perception to training with humans? Addressing this question is accomplished by comparing the \textbf{APPR} and \textbf{HEUR} conditions to \textbf{HUM}.
\end{enumerate}

\subsection{Procedure}

The proposed human-subjects experiment is a single-blind study; waiter participants are not told whether the chef they train with is a human, apprentice or heuristic. To obscure this information, HYBS allows human chefs to play from different real-world locations across a server.
At the start of the experiment, a waiter participant is randomly placed into one of the four conditions (\textbf{HUM, APPR, HEUR, NULL}) and the respective teammate chef will be associated with one of three clusters.
The participant is then given pre-experiment surveys collecting demographic information, familiarity with video games, and opinions on autonomous systems.
Next, each waiter is given a printed slideshow detailing the HYBS game mechanics, the user interface, and teaming objectives.

Waiters assigned to \textbf{HUM}\footnote{\textbf{HUM} condition participants require a human chef from the established chef population as a teammate. This results in a multi-day protocol for human chefs, as established chef training and waiter training are too lengthy to complete within one session.}, \textbf{APPR}, and \textbf{HEUR} conditions undergo a training session while \textbf{NULL} condition waiters do not.  
During training sessions, participants train for 75 minutes with a chef corresponding to their condition and cluster,  with access to video replay of chef gameplay. This video replay---which, in our demonstration, is generated at a rate of two actions per second to obfuscate whether the training session chef is an agent or human---is displayed at the end of each chef performance turn and provides insight to the waiter on their chef teammate's preferred workflow, strategy, and performance. During training, the participant plays at their own pace, allowing for a variable number of games to be completed within the allotted time\footnote{Enforcing a 75-minute limit on waiter training ensured that all necessary data could be collected in a single 120-minute period.
}. 

After training sessions (or lack thereof in the \textbf{NULL} condition), all waiters undergo an evaluation session. Here, waiters are evaluated for three games with each of two human chefs: one who is assigned to the strategy cluster they trained with (in-sample) and another that is not (out-of-sample). During evaluation, the waiter must coordinate with these chefs \emph{without access to video playback} to test whether they have learned a generalizable coordination strategy.
Upon completion of  evaluation sessions, users are provided with surveys gauging team performance with the training session chef and perceived training effectiveness. 
All survey and tutorial materials are provided in this article's supplement.
 
\subsection{Metrics for Cooperative Training Effectiveness} 
\label{sec: metrics}

To measure the effect of a participant's training partner on learning outcomes, we record objective metrics on team performance and evaluation outcomes, as well as subjective scoring by participants of chef partners and training quality.

\subsubsection{Objective Metrics} 
The effectiveness of a waiter-chef team is scored using the total tip (reward) accumulated during training and during evaluation sessions with the out-of-sample chef; values were normalized to range $[0,1]$.
Questions \textbf{Q1}-\textbf{Q3} can be answered using evaluation tip, as this quantity measures the generalizability of a waiter's learned coordination strategy to a partially observable setting with a chef performing an unseen strategy.
On the other hand, cumulative tip during training (in combination with subjective metrics) can be used to answer \textbf{Q4} (Perception) by identifying trends in objective performance of teammates relative to subjective scoring of teammates.  
Our game only records joint teammate performance, and there is no direct measure of individual performance or individual effect on game outcomes. This makes it challenging to infer whether the chef or the waiter contributed more to the overall score (an issue common in teaming environments~\cite{pfau2024real}) or whether individual learning or adaptation has occurred.

\subsubsection{Subjective Metrics} 
To answer \textbf{Q4}, we score the waiter's subjective experience of chefs and training quality using a Likert scale~\cite{batterton2017likert} completed after the evaluation session. The subjective experience of training session chefs is measured using four metrics---the waiter's subjective ratings of trust, reliance, adaptability and predictability of the chef during training---while training quality is aggregated into a single metric. Each subjective metric is obtained through averaging reported alignment with related positive and negative attributes ascribed to a chef or training; see Table \ref{tab: HMT Metric Definitions} for details.

 \begin{table}[t]
     \caption{Attributes associated with subjective metrics on training chef and team performance.  }
     \label{tab: HMT Metric Definitions}
     \scriptsize
  \begin{tabular}{lcc}
    \toprule
         Metric                             &   Positive Attributes           & Negative Attributes  \\
    \midrule
     Trust              &     Likeable, Liked, Trust                &   Deceptive,   
         Suspicious, Underhanded, Wary    \\   \hline
        Reliance          &   Confident, Helpful,  Reliable, Trust    &  Harmful,   No Attention  \\ \hline
        Adaptability       &   Complementary,  Responsive, Sensible     &                           \\ \hline
         Predictability     &   Decreased Workload,  Understood          &   No Sense             \\  \hline
       Training Quality           & Helpful, Important, Preparing, Quality,   Representative & Inadequate,  Insufficient, Poor \\
    \bottomrule
\end{tabular}
\end{table}

\section{Human-Subjects Study Results} \label{sec: human_subject_exp}

We recruited 52 volunteers as subjects in an IRB-approved protocol for this between-subjects study, with ages ranging from 21-58 (mean age: 31.7; standard deviation: 8.8; 43.2\% female)\footnote{The \textbf{NULL} study condition was conducted as a follow-on study post-hoc; all others (\textbf{HUM, APPR, HEUR}) were conducted as a three-condition between-subjects study.
}.
Waiter participants were assigned randomly to different behavior clusters and agent training conditions. Due to higher time demands, established chefs were selected based on schedule availability. 
For each of the collected metrics discussed in Section \ref{sec: metrics}, we tested for normality and equal variance across study conditions and used one-way parametric and nonparametric tests accordingly~\cite{freedman2009statistical}.
Values and the results of statistical analyses on metrics for training and evaluation sessions are visualized in Fig. \ref{fig: metrics}. Details are provided in this article's supplement. 

\subsection{Waiter Training}
\begin{figure}[t]
  \centering
    \includegraphics[width = 0.9\textwidth]{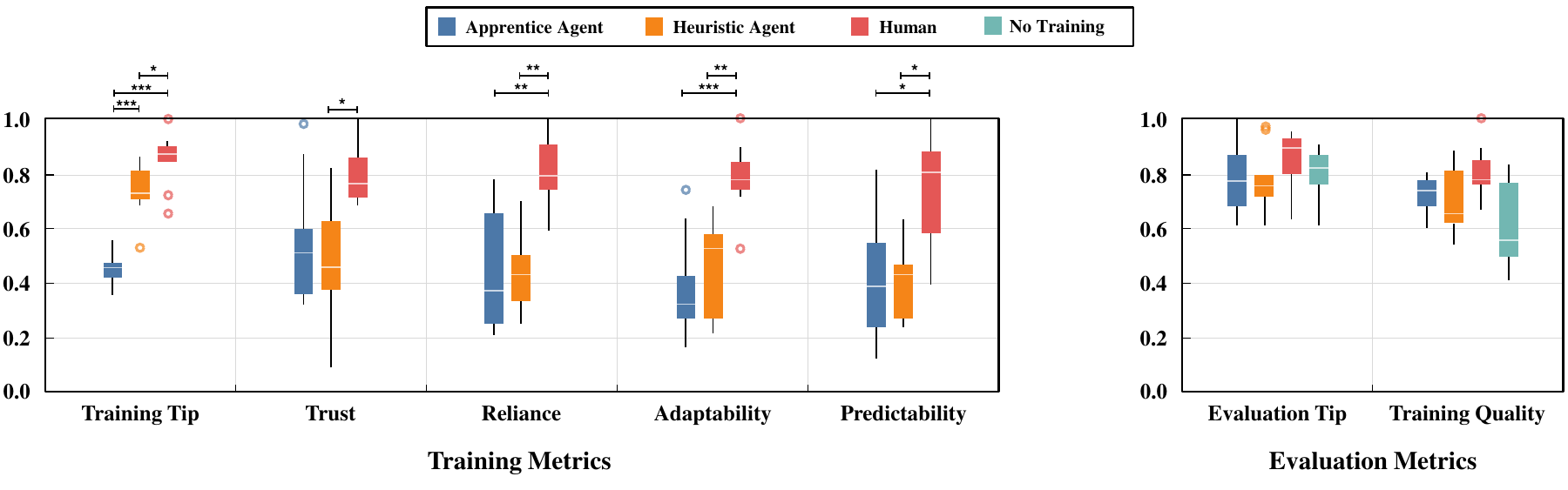}
    \caption{Training and evaluation metrics obtained for the demonstrated human-subjects experiment. Values were normalized to range [0,1], as described in Section \ref{sec: metrics} along with metric definitions. Significant pairwise comparisons are indicated with * if $p\in[0.01,0.05)$,  ** if $p\in[0.001,0.01)$,  and  *** if $p<0.001$. 
    Because evaluation tip in the \textbf{NULL} condition was not significantly different than that of other conditions, we find a negative result for research question \textbf{Q3} (Benefit); thus, this study is not equipped to answer research questions \textbf{Q1} (Effectiveness) or \textbf{Q2} (Personalization). 
    }
    \label{fig: metrics} 
\end{figure}

We find a significant difference in average cumulative tip per game achieved in training sessions across study conditions ($F(2,22) = 55.02$; $p<0.001$), displayed in the leftmost block in Fig. \ref{fig: metrics}. Notably, users trained with human chefs outperformed those trained with apprentice chefs (0.42, [95\% CI 0.32,~0.53]; $p<0.001$) and heuristic chefs (0.13, [0.03,~0.24]; $p<0.05$); users trained with heuristic chefs outperform those trained with apprentice chefs (0.29, [0.19,~0.39]; $p<0.001$). 

Considering Question \textbf{Q4} (Perception), objective metrics on training session chef performance are not entirely predictive of waiters' subjective scoring. Indeed, despite significant differences between \textbf{APPR} and \textbf{HEUR} training tips (0.29, [0.19, 0.39]; $p<0.001$), we find no significant difference in waiters' subjective scoring of agent training chefs across subjective training metrics ($p>0.10$ in all comparisons). If waiters evaluated their chef based on team performance alone, the difference observed in training tip is expected to persist in subjective metrics. Based on the insignificant difference between \textbf{APPR} and \textbf{HEUR} conditions across subjective metrics, we posit that waiters considered expressed behavior in addition to performance when scoring their chef. Indeed, \textbf{HUM} chefs were scored significantly higher than agent chefs across subjective metrics; $p<0.05$ in all comparisons except \textbf{HUM} and \textbf{APPR} on Trust. This suggests that there may other behavioral attributes of a human surrogate model that may need to be improved to ensure positive objective and subjective team collaboration.

\subsection{Waiter Evaluation}
No significant differences were observed in average cumulative tips from evaluation sessions ($F(3,29) = 2.184$; $p>0.100$). 
This is interesting as, during training with complete information, the \textbf{HUM} condition outperformed all others. 
Furthermore, as there is no significant difference between \textbf{NULL} condition evaluation tip and that of other conditions, the effect of training did not significantly impact evaluation outcomes.
This addresses \textbf{Q3} (Benefit), finding that the training conditions designed for this demonstration of an asynchronous cooperative training paradigm did not have a substantial effect on evaluation outcomes. As will be discussed in more detail in Sections \ref{sec: limitations}-\ref{sec: Design Recommendations}, we posit that this could be due to random initialization of evaluation scenarios or that training with full observability did not adequately prepare users for evaluation in a partially observable environment. 
Due to the disparity in findings across training and evaluation, this demonstration is unable to definitively answer \textbf{Q1} (Effectiveness) and \textbf{Q2} (Personalization).  

Differences in Training Quality are trending ($F(3,29) = 2.184$; $p=0.11$). Although we expected the \textbf{NULL} condition would rate training as inadequate (as they only received a slideshow), our experiment is between-subjects and these participants may not have had a good point of comparison for this metric, possibly explaining the high variance in \textbf{NULL} participants' scoring of Training Quality.

\subsection{Limitations and Future Work}
\label{sec: limitations}

This article provides a feasible approach to evaluating cooperative asynchronous training systems. As this evaluation is novel, we evaluate our study in hindsight and present key insights based on observed outcomes that would have improved our ability to answer our research questions. Future work should consider these limitations when designing subsequent human-subjects experiments for similar problems. 

\begin{enumerate}
    \item \textbf{Untangling individual capability effects from team cohesiveness effects.} Although it was possible to measure the performance of a team in HYBS, it was difficult to understand whether team performance was due to high/low-performing individual teammates or poor team cohesiveness. In the future, it would be beneficial to design surrogate measures that provide insight into individual behaviors---such as \textit{obedience} (i.e., how frequently a chef follows the waiter's recommendations)---which may enable some understanding of a chef's use of communicated information from the waiter. However, even so, it remains difficult to measure waiter performance holistically in teaming environments~\cite{pfau2024real}. This is especially true for environments like HYBS in which teammates operate under very different roles. 
    Future work should investigate new metrics that may enable teasing out teammate performance.

    \item \textbf{Waiter training and evaluation sessions were not identical.} The waiter training provided access to video replay, better allowing waiters to understand the decisions chefs made during gameplay (full observability). The waiter evaluation removed video replay to test whether users were able to function well under partial observability and learn a generalized strategy. While utilizing full observability during training may have been beneficial in improving training speed and quality, this may have not adequately prepared users for teaming under partial observability. Future work should explore techniques that minimize training and testing mismatch while ensuring high user training speed.
    
    \item \textbf{Low sample size:} 
    The implemented human-subjects experiment had a relatively low sample size, which may have contributed to a lack of significance in our results for \textbf{Q3} and prohibited meaningful intra-condition post-hoc analyses. 
    
\end{enumerate}

\subsection{Design Recommendations for Cooperative Asynchronous Training Systems} \label{sec: Design Recommendations}

Significant research is needed to fully develop a cooperative asynchronous training paradigm using automated agents. Based on the demonstrated human-subjects experiment discussed in Section \ref{sec: human_subject_exp}, we provide the following design recommendations for future work in building such a system:

\begin{enumerate}

    \item \textbf{Behavior clustering enables tractable behavior-based study and statistical analysis.}
    When stratifying a data collection by human demonstrator is infeasible due to a large number of contributors, a tractable experiment can be segmented on behaviors derived from unsupervised clustering~\cite{franceschi2019convolutionalencoder} of demonstration data. 
    Specifically, categorizing by behavior clusters is expected to enable controlling for demonstrator behaviors (rather than for individual demonstrators) in a two-way statistical test~\cite{freedman2009statistical}, substantially reducing the factorial requirements for a comparable two-factor study~\cite{bailey2008design}. This, in turn, is expected to substantially reduce the number of participants needed for a two-factor human-subjects experiment, and eases scheduling requirements, as a single user is not expected to be an evaluation partner for all teammate participants.

    \item \textbf{Training and evaluation sessions should be carefully designed to focus on teaming desiderata.} Training sessions did not have as strong an effect as anticipated, resulting in little difference between \textbf{NULL} and other study conditions in evaluation outcomes.  This may be due to our use of fully randomized scenarios in our demonstration. 
    Although randomized scenarios reduced experimenter bias, it may have also dampened the importance of training sessions. 
    With more deliberate scenario design (e.g., of customer profiles and potato ingredients) tailored toward enforcing teammates' reliance on one another, it may be possible to reduce noise in our data collection and ensure a need for teaming.
    
    \item \textbf{Add proficiency checks to ensure training is sufficient.} As we could not directly track individual waiter performance, we could not measure whether waiter performance had peaked or settled, preventing us from knowing whether training sessions were long enough for learning to complete. Future work should utilize quantitative and qualitative assessments of training sufficiency to determine whether users' training time is sufficient for learning. This may be challenging, as users may have different rates of learning and require variable time.
    
    \item \textbf{Identify automated agent qualities that lead to positive training experiences.}
    Despite significant differences in training tip (an objective metric on chef performance), no significant differences were observed in subjective scoring of the evaluated automated agents across subjective metrics. This result aligns with others in the literature~\cite{siu2021evaluation, zhang2021ideal}: that agent performance may be less indicative of a human teammate's acceptance than behavior. We see improving autonomous agent behavior to align with human expectations 
    as an essential element of future cooperative asynchronous training paradigms.
     
    \item \textbf{Identify performant agents \textit{a priori}. } 
    Evaluating both subjective and objective agent efficacy prior to running the experiment is expected to reduce risk and ultimately provide a better user experience to participants teaming with those agents. For example, established chefs could interact with a suite of agent chefs trained on their demonstrations and rate their performance to inform which agents are included in later stages of the experiment.

\end{enumerate}

\section{Conclusions}\label{sec: Conclusion}

A cooperative asynchronous training paradigm that allows autonomously-controlled robots to act as surrogate teammates in a partially-observable coordination setting is expected to reduce time constraints~\cite{gerencser2020review, marano2020review, serna2016teaching} when scheduling human teammates is not possible or cost-prohibitive, but where building a mental model of one's teammate is essential for success in evaluation~\cite{andrews2023role, Paleja2022TheUO}. To test the efficacy of autonomous robots as training partners in cooperative asynchronous training, this study introduces a novel experimental design in which humans train with human and automated agents on complex tasks in a limited communication, partially observable environment. The experimental design addresses a key question underlying agent efficacy: whether training with an agent yields comparable or superior efficacy to training with a human, given that training itself is effective.
While the findings of our implementation were mixed, we believe that this experimental design and the lessons learned from it provide an important contribution towards development of cooperative asynchronous training.

\bibliographystyle{unsrt}
\bibliography{main_arXiv}

\end{document}


\maketitle

\blfootnote{\newline \noindent DISTRIBUTION STATEMENT A. Approved for public release. Distribution is unlimited.

\noindent This material is based upon work supported by the Under Secretary of Defense for Research and Engineering under Air Force Contract No. FA8702-15-D-0001. Any opinions, findings, conclusions or recommendations expressed in this material are those of the author(s) and do not necessarily reflect the views of the Under Secretary of Defense for Research and Engineering.} 

This supplement is organized as follows. Section \ref{supp: overcooked} provides details on the Overcooked-AI: Have You Been Served?  (HYBS) teaming game created for this study, including the process for randomizing scenarios in the resulting demonstrated human-subjects experiment. Section \ref{supp:seq_enc} details the state feature extraction used for unsupervised behavior inference mechanism used to partition established chefs into groups expressing similar behaviors. Section \ref{supp: apprentice} and \ref{supp: heuristic} detail apprentice and heuristic agent chef generation, respectively. Section \ref{supp: statistics} contains details on the statistical analysis presented in Section 6 of the original paper. 
\section{Details on Overcooked-AI: Have You Been Served? Edition Teaming Game} \label{supp: overcooked}

\begin{figure}[t]
    \centering
    \subfigure[Example Waiter Gameplay]{\includegraphics[width = 0.48\textwidth]{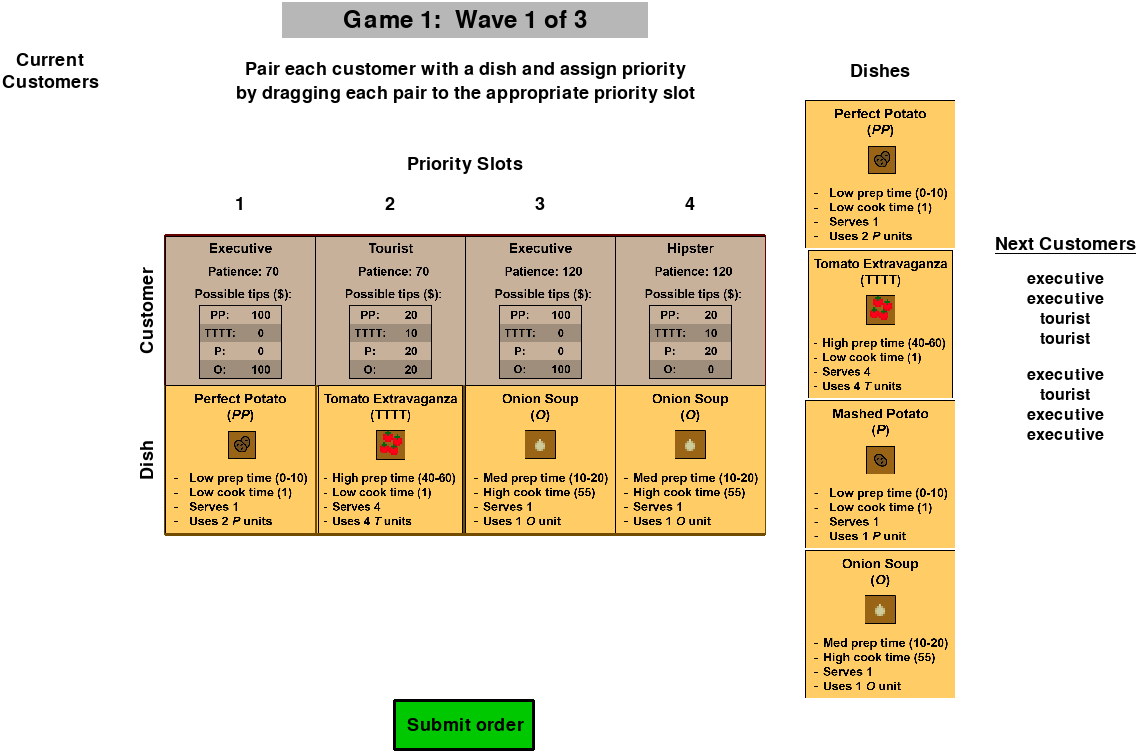}} 
    \subfigure[Example Chef Gameplay]{\includegraphics[width = 0.48 \textwidth]{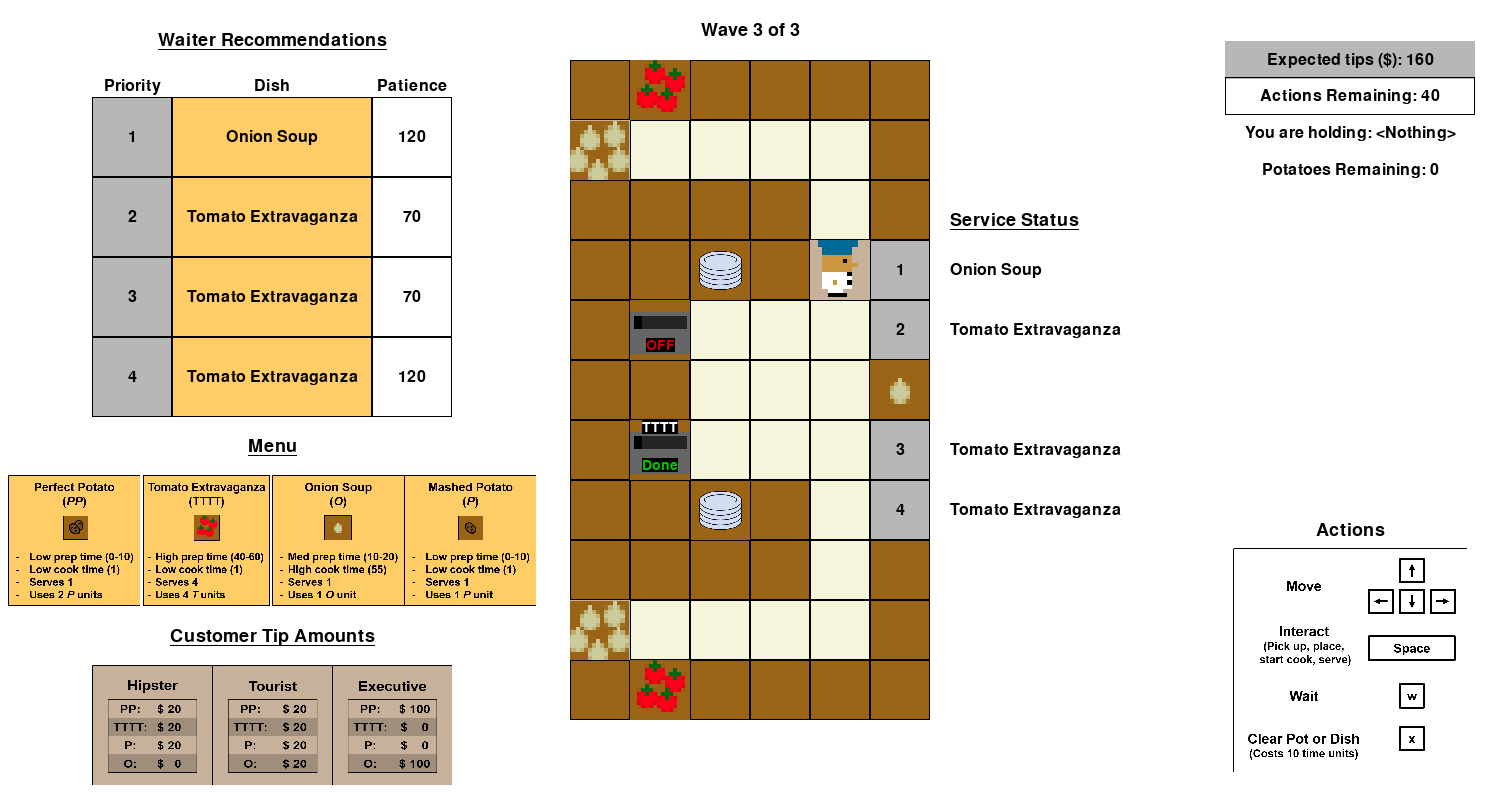}}
    \caption{Visualization of the gameplay in the two roles (waiter and chef) in the modified version of Overcooked-AI~\cite{carroll2019utility} considered in this article. }
    \label{fig: Overcooked}
\end{figure}

This section provides details on HYBS, the gameplay of which is visualized in Figure \ref{fig: Overcooked}. In a round of HYBS---which was derived from Overcooked-AI~\cite{carroll2019utility}---the chef must navigate the kitchen to obtain ingredients, place them in a pot, and serve those ingredients to customers. Some ingredients (tomato and onion) have an unlimited inventory, while another (potato) is designed to be highly-effective but is limited in number. Importantly, onion and tomato ingredients must be collected from a region far from pots (making collection costly) while potato ingredients can be accessed nearby pots. The chef is given a limited number of steps or ``action points'' to collect ingredients and serve dishes, so they must plan realistic servings of customers based on waiter recommendations. Action points are deducted for moving, placing or picking up an object, active waiting, or clearing a dish from the pot. In the initial preparation round, the chef is given 15 action points to stage ingredients; in regular rounds, the chef is given 135 action points. 

Chefs are capable of serving four dishes to customers; different dishes require different ingredient combinations and cooking time and are designed to be applicable to different sets of customers in varying scenarios. 
Potato-based dishes---Perfect Potato (\textbf{PP}) and Mashed Potato (\textbf{P})---can be quickly cooked with little preparation time (the time needed to gather ingredients) in order to serve a potentially high-value customer; \textbf{PP} (which requires 2 potato ingredients) tends to correspond to higher tips than \textbf{P} (which only requires one potato ingredient) at the cost of more resource usage. 
Onion Soup (\textbf{O}) can be prepared using only one onion ingredient, but it has a longer cook time (55 action points). Although Tomato Extravaganza (\textbf{TTTT}) has higher preparation time (requiring four tomato ingredients), it produces four servings that may be provided to customers in a round or staged for future rounds. All other dishes and the pots are cleared between rounds. Importantly, unlike potato-based dishes, the ingredients required for \textbf{O} and \textbf{TTTT} are unlimited. Dishes in HYBS are overviewed in Table \ref{tab:dishes}. 

\begin{table}[b]
    \centering
    \caption{Requirements and outcomes of dishes in HYBS.}
    \footnotesize
    \begin{tabular}{cccc}
    \toprule
        Dish                &   Ingredients                 &   Cook Time                   &   Servings    \\ \midrule
        \textbf{PP}: Perfect Potato      &   2 Potatoes                  &   1 Action                    &   1 Customer  \\ \hline
        \textbf{P}: Mashed Potato       &   1 Potato                    &   1 Action                    &   1 Customer  \\ \hline
        \textbf{O}: Onion Soup         &   1 Onion                     &   55 Actions                  &   1 Customer  \\ \hline
        \textbf{TTTT}: Tomato Extravaganza             &   4 Tomatoes &    1 Action  &  4 Customers  \\  
        \bottomrule
    \end{tabular}
    \label{tab:dishes}
\end{table}
 
For the demonstrated experiment, we built three customer profiles with varying dish preferences (and resulting tips) that must be managed by a chef-waiter team. 
First, Executive customers provide high rewards (\$100)  for \textbf{PP} or \textbf{O} dishes, but yield no tips for \textbf{P} or \textbf{TTTT} dishes. 
Second, Hipster customers give a low but equal tip (\$20) for \textbf{PP} and \textbf{P} dishes, a lower tip (\$10) for \textbf{TTTT}, and no tip for \textbf{O} dishes.
Finally, Tourist customers give a low but equal tip (\$20) for \textbf{PP}, \textbf{P}, and \textbf{O} dishes and a lower tip (\$10) for \textbf{TTTT}. 
Because the chef is only aware of customer-dish pairs (not customer profiles), they must follow waiter recommendations subject to resource constraints. Despite high rewards for potato-based ingredients, incorrect usage of these limited ingredients could yield suboptimal tips. For example,  giving a Hipster or Tourist a \textbf{PP} dish would be a waste of resources, as a \textbf{P} dish would yield the same tip. On the other hand, attempting to save resources by giving an Executive a \textbf{P} dish will result in waste of 1 potato ingredient, as that serve event will yield \$0 tip. As such, chefs were incentivized to follow waiter recommendations, subject to resource availability. Customer profiles are summarized in Table \ref{tab:customers}.

\begin{table}[t]
    \centering
    \caption{Customer Profiles in HYBS. }
    \footnotesize
    \begin{tabular}{ccccc}
    \toprule
        Customer    &   \textbf{PP} Tip  &   \textbf{P} Tip   &   \textbf{O} Tip   &   \textbf{TTTT} Tip    \\ \midrule 
        Executive   &   100     &   0       &   100     &   0           \\ \hline
        Hipster     &   20      &   20      &   0       &   10          \\ \hline
        Tourist     &   20      &   20      &   20      &   10          \\ \bottomrule
    \end{tabular}
    \label{tab:customers}
\end{table}

Finally, each customer is randomly assigned a ``patience'' score that is associated with how long the chef has to serve the customer until they leave the kitchen and become unservable. The clock starts at the beginning of a chef performance round and ticks down by 1 for every action point used by the chef. A customer becomes unservable once the number of action points taken by the chef exceeds the customer's patience score; e.g., a chef that took 71 actions to serve a customer with a patience of 70 would receive no tip for that serve event. 

Each participant of our human-subjects experiment played the same set of random scenarios. The parameters varying across random scenarios consisted of 1) the number of available potato ingredients; 2) the customers to be served; and 3) customer patience scores. Scenarios are generated by uniformly sampling a number of potatoes from \{1, 2, 3, 4, 5\}, customer profiles from the set described in Table \ref{tab:customers}, and patience scores in \{70, 120\}.

\section{Data Used for Sequential Encoding} 
\label{supp:seq_enc}

This section describes the state variables used for the behavior encoding step discussed in Section 4.1 of the main article. As referenced in the main text, we selected state variables that provide a meaningful representation of the game at the time a user selected an action. The specific state variables used are summarized in Table \ref{tab: State-action pairs}. 


\begin{table}[b]
    \centering
    \caption{Description of state-action pairs used for sequential encoding of human action sequences in this study's established chefs. } 
    \footnotesize
    \begin{tabular}{cccc}
    \toprule
        Variable                      &   Outcomes                                          \\ \midrule
      Action taken      &   Move, Interact, Clear, Wait               \\ \hline
      Chef-held object  &   Nothing, Ingredient, Dish, Plate     \\  \hline
      Customers served  &   Subset of Customers 1-4     \\  \hline
        Waiter-assigned dishes    & Assigned Dishes                                  \\  \bottomrule
    \end{tabular}
    \label{tab: State-action pairs}
\end{table}

\section{Apprentice Agent Training} \label{supp: apprentice}

Apprentice agent generation maintained three steps, 1) Data Labeling, 2) Personalized Apprenticeship Learning utilizing the framework of \cite{Paleja2020InterpretableAP}, and 3) Personalized Embedding Storage. 

\textbf{Data Labeling --} 
To enable apprentice agent training~\cite{Paleja2020InterpretableAP} in support of the \textbf{APPR} condition, we randomly sample gameplay from each derived cluster in the obtained clustering discussed in this section. The smallest cluster (Cluster 1) consists of only 27 games. Thus, the training set consists of all 27 games from cluster 1 and a uniformly random sample of 27 games from Clusters 2 and 3. As such, each identified behavior has equal representation in the training set of apprentice models.  
To encode a meaningful representation of the world, we selected the following state variables: the current location of the chef, object placements, waiter-recommended dishes, served dishes so far, and what object is held by the chef (if any). We identified high-level actions (e.g., get an onion or stage a tomato) associated with states at a given time step.
The benefit of utilizing a high-level action space that predicts goals as opposed to cardinal actions is that the apprenticeship learning problem is easier, and we are able to utilize a goal-oriented action planner to plan low-level actions; see Section \ref{supp: heuristic} for details. Indeed, directly creating agents that function on cardinal actions may lead to noisy and nonsensical behaviors in regions of the high-dimensional state space, leading to less proficient agents that are easily detected as non-human players. 

\textbf{Personalized Apprenticeship Learning --} Once we generated a set of state-goal pairs for each user's set of games, we begin the process of apprenticeship learning. We start by uniformly initializing a personalized embedding for each user, a neural network which predicts a goal conditioned on a user's state and embedding, and a network that recovers the embedding given a state and goal. 
Throughout training, we sampled data from a respective user, loaded in the personalized embedding, and updated 1) the neural network parameters and personalized embedding via a cross-entropy loss that maximizes imitation performance, and 2) updated all parameters utilizing a mutual-information maximization loss that maximizes the relation between sampled embeddings and state-goal pairs (i.e., mode discovery). We refer users to \cite{Paleja2020InterpretableAP} for specific details regarding the training procedure.

\textbf{Personalized Embedding Storage --} Upon training completion, we stored a personalized embedding for each human chef. These embeddings are three-dimensional and capture stylized behaviors across users. For each strategy cluster produced via Section \ref{supp:seq_enc}, we utilized the mean personalized embedding across all users assigned to that cluster to generate an apprentice chef associated with that cluster.

It is important to note that the apprentice predicts a single goal, which may make it difficult to imitate users that pursue multiple goals at once. However, unlike the heuristic, the apprentice is free to deviate from waiter-assigned dishes, allowing for the ability to personalize to users that acted in specialized ways.

\section{Heuristic Agent Description} \label{supp: heuristic}

Our goal for designing the heuristic agent is to create a teammate decision process that would be logical, performant, and---as much as possible---predictable. To this end, we combine a heuristic-aided path search for raw game actions with a simple goal selection algorithm. Due to limited compute time available for planning during our experiment, we also augment the game action search with a another search over higher-level actions.

\textbf{Game Action Search --} Our heuristic agents rely on Goal-Oriented Action Planning (GOAP)~\cite{orkin2004goap} to determine the fewest actions needed to achieve a goal. GOAP uses actions and constraints to enumerate viable chains of actions and the A* search algorithm to discover the minimum-cost action plan. Within the HYBS environment, plans for simple goals can be computed in much less than a second. However, for more complex goals ("gather, prepare, and serve onion to Customer A", "serve tomato to Customer B and potato to Customer C"), the run time exceeds the wait time for the waiter teammate. This requires us to combine GOAP with a higher-level planning scheme.

\textbf{High-level Action Search --}
To reduce run time for complex tasks, we augment the GOAP planner with a search scheme over higher-level ("macro-") actions. The high-level planner computes a chain of macro-actions ("get potato", "place in pot", "cook potato", "serve potato to Customer A") which generates a chain of goals that are passed sequentially to the GOAP planner. We choose Monte Carlo Tree Search~\cite{kocsis2006bandit}---an anytime algorithm~\cite{hendler2014artificial} that supports large decision space exploration---for our macro-action planning. We limit the high-level search to ten thousand iterations based on time restrictions from the experiment. We note also that the faster decision time provided by the macro-action search comes with the cost of omitting potential solutions that require interrupting a macro-action in the middle of its execution in pursuit of better performance via multi-tasking.

\textbf{Goal selection --} To achieve our logic and predictability goals, we program the agent to be as obedient as possible to waiter recommendations and select goals for the heuristic agent using the following steps. We start by planning a path to the highest-priority waiter recommendation. If successful, we combine the next-highest priority recommendation in the waiter list with the previous and re-plan; otherwise, we move to the next service goal and re-plan. If the combined goal cannot be reached, we omit the newly-added goal from future consideration and proceed to the next. If the combined goal is reached, we add the next goal in the waiter list and repeat the process above until no more service goals can be met and the waiter list is exhausted. If any time remains after planning the service goals, we attempt to combine ingredient staging with the service goals and repeat the planning process until the game time budget is nearly exhausted. The final few game steps are used to return the agent to the center of the board.

To avoid exhausting the patience of the waiter teammate during the experiment, we terminate action planning after a few minutes of search time. For most cases this is sufficient to achieve the highest-priority goals; however, occasionally some goals theoretically achievable by the agent are only partially fulfilled.

\section{Statistical Analysis} \label{supp: statistics}

This section contains details for the statistical analyses that are contained in Section 6~\cite{freedman2009statistical}. For each metric in Section 5.2, we perform the following statistical tests in $R$: 

\begin{enumerate}
    \item A Shapiro-Wilk (SW) normality test to test the hypothesis that metric values are distributed normally.
    \item A Bartlett test of homogeneity of variances to test the hypothesis of equal variance for metric values across treatment groups.  
\end{enumerate}

The results of this statistical analysis are provided in Table \ref{tab: normality}. If the $p$-values resulting from each of these statistical tests is above the $\alpha=0.05$ significance level, then the assumptions of a one-way ANOVA are met. If at least one of the above $p$-values is above the $\alpha=0.05$ significance threshold, we had planned to resort to non-parametric statistical tests: a Kruskal-Wallis test with the Wilcoxon rank sum test with Bonerroni correction for multiple hypothesis testing for pairwise comparisons. As is shown in Table \ref{tab: normality}, all metrics satisfy the assumptions of the one-way ANOVA test. Thus, we use the parametric one-way ANOVA to test whether metric values are significantly different across treatments. We perform post-hoc pairwise comparisons using the Tukey-Cramer honest significance test.  

\begin{table}[t]
    \centering 
    \footnotesize
   \begin{tabular}{ccc}
    \hline
        Metric          &   SW $p$-value                        &   Bartlett  $p$-value     \\ \hline
        Training Tip    &   $5.82\times 10^{-2}$                &   $2.76\times 10^{-1}$    \\\hline
        Trust           &   $5.96\times 10^{-1}$                &   $1.26\times 10^{-1}$                  \\\hline
        Reliance        &   $1.26\times 10^{-1}$                &   $3.59\times 10^{-1}$                  \\\hline
        Adaptability    &   $3.86\times 10^{-1}$                &   $7.41\times 10^{-1}$                  \\\hline
        Predictability  &   $1.53\times 10^{-1}$                &   $3.66\times 10^{-1}$                  \\\hline
        Evaluation Tip(Seen)  &   $1.53\times 10^{-1}$                &   $3.66\times 10^{-1}$                  \\\hline
        Evaluation Tip(Unseen)  &   $1.04\times 10^{-1}$                &   $8.92\times 10^{-1}$                  \\\hline
        Training Quality   &  $6.87\times 10^{-1}$        &   $6.88\times 10^{-1}$                  \\ \hline
    \end{tabular}
    \caption{Results of tests for normality. Insignificance to the $\alpha=0.05$ significance level indicates applicability of the parametric one-way ANOVA test. Notably, all metrics meet the assumptions of one-way ANOVA.}
    \label{tab: normality}
\end{table}

Results of ANOVA statistical testing are provided in Table \ref{tab: anovas}. All statistical tests achieve significance at the $\alpha=0.05$ significance level except for the ones performed on evaluation metrics: Evaluation Tip (Seen and Unseen) and Training Quality. We provide statistics in Table \ref{tab: pairwise comparisons} on pairwise comparisons obtained with Tukey-Cramer Honest Significance Test. We find significant differences between study conditions' training tip for each pairwise comparison ($p<0.05$) but find no significant difference in any pairwise comparison in evaluation tip ($p>0.10$). This includes insignificance in evaluation tip comparisons between the \textbf{NULL} condition (which has no training condition) and any other treatment group (which has a training condition). We find significant differences between subjective metrics related to training (Trust, Reliance, Adaptability, and Predictability) comparing human chefs to agent chefs (\textbf{APPR} or \textbf{HEUR}). However, we find no significant difference between \textbf{APPR} and \textbf{HEUR} scores on these metrics. Finally, for the Training Quality metric, no significant difference is observed between treatments, albeit that the comparison between \textbf{NULL} and \textbf{HUM} Training Quality is trending ($p=0.072$).

\begin{table*}[b]
    \centering 
    \footnotesize
    \begin{tabular}{ccccccc}
    \toprule
        \multicolumn{2}{c}{Metric}                &   Degrees of Freedom      &   Sum Square  &   Mean Square &   $F$-Value  &   $p$-value               \\ \midrule 
        \multirow{2}*{Training Tip}& Treatment      &   2                       &   0.761       &   0.381       &   55.020     &    $\mathbf{2.75\times 10^{-9}}$    \\ 
                                & Residual          &   22                      &   0.152       &   0.007       &   --         &   --                      \\ \hline 
        \multirow{2}*{Trust}    & Treatment         &   2                       &   0.467       &   0.234       &   4.868       &    $\mathbf{1.78\times 10^{-2}}$    \\ 
                                & Residual          &   22                      &   1.055       &  	0.048       &   --         &   --                      \\ \hline 
        \multirow{2}*{Reliance} & Treatment         &   2                       &   0.663       &   0.332       &   9.305      &    $\mathbf{1.18\times 10^{-3}}$    \\ 
                                & Residual          &   22                      &   0.784       &   0.036       &   --         &   --                      \\ \hline 
        \multirow{2}*{Adaptability} & Treatment     &   2                       &   0.736       &   0.368       &   11.160     &    $\mathbf{4.51\times 10^{-4}}$    \\ 
                                & Residual          &   22                      &   0.725       &   0.033       &   --         &   --                      \\ \hline 
        \multirow{2}*{Predictability} & Treatment   &   2                       &   0.536       &   0.268       &   6.320      &    $\mathbf{6.78\times 10^{-3}}$    \\ 
                                & Residual          &   22                      &   0.933       &   0.042       &   --         &   --                      \\ \hline 
        \multirow{2}*{\shortstack[c]{Evaluation Tip\\ (\textit{Seen})}}& Treatment    &   3     &   0.031       &   0.010       &   0.698      &   $5.61\times 10^{-1}$    \\ 
                             & Residual             &   29                      &   0.428       &   0.015       &   --         &   --                      \\ \hline 
        \multirow{2}*{\shortstack[c]{Evaluation Tip\\ (\textit{Unseen})}}& Treatment    &   3                       &   0.016       &   0.005       &   0.714      &   $5.51\times 10^{-1}$    \\ 
                             & Residual             &   29                      &   0.219       &   0.008       &   --         &   --                      \\ \hline 
        \multirow{2}*{Training Quality}       & Treatment     &   3       &   0.145       &   0.048       &   2.184      &   $1.11\times 10^{-1}$    \\ 
                                                    & Residual      &   29      &   0.641       &   0.022       &   --         &   --                      \\ \bottomrule 
    \end{tabular}
    \caption{Results of ANOVA on metric values from the demonstrated human subject experiment. $p$-values significant to the $\alpha= 0.05$ significance level are indicated in bold. All ANOVAs except for that of Evaluation Tip are significant to the $\alpha=0.05$ significance level. }
    \label{tab: anovas}
\end{table*}

\begin{table*}[t]
    \centering 
    \footnotesize
    \begin{tabular}{ccccc}
    \toprule
        Metric                          &   Comparison                  &   Difference, [95\% CI]   &   $p$-value                  \\ \midrule 
        \multirow{3}*{Training Tip}     &   \textbf{HEUR} -- \textbf{APPR}     &   0.289, [0.190, 0.387]   &   $\mathbf{6.57\times 10^{-7}}$    \\
                                        &   \textbf{HUM} -- \textbf{APPR}         &   0.421, [0.316, 0.526]   &   $\mathbf{3.27\times 10^{-9}}$    \\
                                        &   \textbf{HUM} -- \textbf{HEUR}          &   0.132, [0.027, 0.237]   &   $\mathbf{1.23\times 10^{-2}}$    \\ \hline 
        \multirow{3}*{Trust}            &   \textbf{HEUR} -- \textbf{APPR}     &   -0.069, [-0.329, 0.190] &   $7.82\times 10^{-1}$    \\
                                        &   \textbf{HUM} -- \textbf{APPR}         &   0.263, [-0.015, 0.540]  &   $6.58\times 10^{-2}$    \\
                                        &   \textbf{HUM} -- \textbf{HEUR}          &   0.332, [0.055, 0.609]   &    $\mathbf{1.71\times 10^{-2}}$   \\ \hline 
        \multirow{3}*{Reliance}         &   \textbf{HEUR} -- \textbf{APPR}     &   0.007, [-0.217, 0.230]  &   $9.97\times 10^{-1}$      \\
                                        &   \textbf{HUM} -- \textbf{APPR}         &   0.366, [0.127, 0.605]   &    $\mathbf{2.42\times 10^{-3}}$     \\
                                        &   \textbf{HUM} -- \textbf{HEUR}          &   0.359, [0.120, 0.598]   &    $\mathbf{2.86\times 10^{-3}}$    \\ \hline 
        \multirow{3}*{Adaptability}     &   \textbf{HEUR} -- \textbf{APPR}     &   0.064, [-0.151, 0.279]  &   $7.36\times 10^{-1}$     \\
                                        &   \textbf{HUM} -- \textbf{APPR}         &   0.409, [0.180, 0.639]   &    $\mathbf{5.35\times 10^{-4}}$     \\
                                        &   \textbf{HUM} -- \textbf{HEUR}          &   0.345, [0.115, 0.575]   &    $\mathbf{2.90\times 10^{-3}}$    \\ \hline 
        \multirow{3}*{Predictability}   &   \textbf{HEUR} -- \textbf{APPR}     &   -0.009, [-0.252, 0.235] &   $9.96\times 10^{-1}$     \\
                                        &   \textbf{HUM} -- \textbf{APPR}         &   0.322, [0.061, 0.582]   &    $\mathbf{1.39\times 10^{-2}}$     \\
                                        &   \textbf{HUM} -- \textbf{HEUR}          &   0.330, [0.070, 0.591]   &    $\mathbf{1.15\times 10^{-2}}$    \\ \hline  
        \multirow{6}*{\shortstack[c]{Evaluation Tip\\ (\textit{Seen})}}   &   \textbf{HEUR} -- \textbf{APPR}     &   0.002, [-0.154, 0.158]  &   $9.99\times 10^{-1}$    \\
                                        &   \textbf{HUM} -- \textbf{APPR}         &   0.077, [-0.090, 0.244]  &   $5.95\times 10^{-1}$    \\
                                        &   \textbf{HUM} -- \textbf{HEUR}          &   0.075, [-0.091, 0.242]  &   $6.11\times 10^{-1}$    \\  
                                        &   \textbf{NULL} -- \textbf{APPR}          &   0.038, [-0.123, 0.199]  &   $9.15\times 10^{-1}$    \\
                                        &   \textbf{NULL} -- \textbf{HEUR}           &   0.037, [-0.124, 0.197]  &   $9.25\times 10^{-1}$    \\
                                        &   \textbf{NULL} -- \textbf{HUM}               &   -0.039, [-0.210, 0.132] &   $9.26\times 10^{-1}$    \\ \hline
        \multirow{6}*{\shortstack[c]{Evaluation Tip\\ (\textit{Unseen})}}   &   \textbf{HEUR} -- \textbf{APPR}     &   -0.035, [-0.147, 0.0768]  &   $8.29\times 10^{-1}$    \\
                                        &   \textbf{HUM} -- \textbf{APPR}         &   0.029, [-0.090, 0.148]  &   $9.14\times 10^{-1}$    \\
                                        &   \textbf{HUM} -- \textbf{HEUR}          &   0.063, [-0.056, 0.183]  &   $4.81\times 10^{-1}$    \\  
                                        &   \textbf{NULL} -- \textbf{APPR}          &   -0.002, [-0.117, 0.113]  &   $9.99\times 10^{-1}$    \\
                                        &   \textbf{NULL} -- \textbf{HEUR}           &   0.033, [-0.082, 0.148]  &   $8.61\times 10^{-1}$    \\
                                        &   \textbf{NULL} -- \textbf{HUM}               &   -0.030, [-0.153, 0.092] &   $9.06\times 10^{-1}$    \\ \hline
        \multirow{6}*{Training Quality}&   \textbf{HEUR} -- \textbf{APPR}&   0.001, [-0.190, 0.191]  &   $9.99\times 10^{-1}$    \\
                                        &   \textbf{HUM} -- \textbf{APPR}         &   0.098, [-0.106, 0.303]  &   $5.61\times 10^{-1}$    \\
                                        &   \textbf{HUM} -- \textbf{HEUR}          &   0.098, [-0.106, 0.302]  &   $5.66\times 10^{-1}$    \\
                                        &   \textbf{NULL} -- \textbf{APPR}          &   -0.098, [-0.294, 0.098]  &   $5.32\times 10^{-1}$   \\
                                        &   \textbf{NULL} -- \textbf{HEUR}           &   -0.099, [-0.296, 0.098] &   $5.29\times 10^{-1}$    \\
                                        &   \textbf{NULL} -- \textbf{HUM}               &   -0.197, [-0.406, 0.013]&   $7.20\times 10^{-2}$    \\ \bottomrule 
    \end{tabular}
    \caption{Results of post-hoc comparisons. Differences with 95\% confidence intervals (CIs) are shown for pairwise comparisons made with Tukey-Cramer's honest significance test. $p$-values significant to the $\alpha= 0.05$ significance level are indicated in bold.  }
    \label{tab: pairwise comparisons}
\end{table*}

\bibliographystyle{unsrt}
\bibliography{main_arXiv}